\documentclass{ar-1col-S2O}
\usepackage[numbers]{natbib}
\usepackage{url}
\usepackage{amsmath,amssymb, amsthm}
\usepackage{bm}
\usepackage{physics}

\jname{Xxxx. Xxx. Xxx. Xxx.}
\jvol{AA}
\jyear{YYYY}
\doi{10.1146/((please add article doi))}

\begin{document}

\markboth{M. M. Telo da Gama and R. C. V. Coelho}{Phase Separation in Mixtures of Nematic and Isotropic Fluids}

\title{Phase Separation in Mixtures of Nematic and Isotropic Fluids}

\author{Margarida M. Telo da Gama$^{1,2,3}$ and Rodrigo C. V. Coelho$^{1,4}$
\affil{$^1$Centro de Física Teórica e Computacional, Faculdade de Ciências, Universidade de Lisboa, 1749-016 Lisboa, Portugal.; email: mmgama@ciencias.ulisboa.p}
\affil{$^2$Departamento de Física, Faculdade de Ciências,
Universidade de Lisboa, P-1749-016 Lisboa, Portugal.}
\affil{$^3$International Institute for Sustainability with Knotted Chiral Meta Matter (WPI-SKCM$^2$), Hiroshima University, Higashi-Hiroshima, Hiroshima 739-8526, Japan.}
\affil{$^4$Centro Brasileiro de Pesquisas Físicas, Rua Xavier Sigaud 150, 22290-180 Rio de Janeiro, Brazil}
}

\begin{abstract}
Mixtures of nematic liquid crystals and isotropic fluids display a diverse range of phase behaviors, arising from the coupling between orientational order and concentration fluctuations. In this review, we introduce a simplified mathematical framework that integrates the Landau–de Gennes free energy for nematic ordering with the Cahn–Hilliard free energy for phase separation. We derive the corresponding governing equations and analyze the stability of uniform phases, along with the resulting interfacial phenomena. The review concludes with a brief discussion highlighting key differences in phase separation between mixtures of isotropic fluids with passive and active nematics. 
\end{abstract}

\maketitle

\tableofcontents

\section{Introduction}

Liquid crystals (LCs) are a unique state of matter that combines the fluidity of isotropic liquids with the anisotropic properties of crystalline solids \cite{deGennes1993}. Unlike conventional phases—solid, liquid, and gas—liquid crystals exist in an intermediate state where molecules retain some degree of order, similar to a solid, while maintaining the ability to flow like a liquid. This combination enables liquid crystals to respond dynamically to external stimuli such as temperature, electric fields, and pressure, making them highly versatile for technological applications. LCs are categorized into two major types based on phase behavior and the conditions that promote order: thermotropic and lyotropic \cite{deGennes1993,Collings2019}. Thermotropic LCs primarily respond to temperature variations, whereas lyotropic LCs form in the presence of a solvent.

Thermotropic LCs undergo temperature-driven phase transitions and are classified into subtypes based on the degree of molecular order, with nematic (N) being the most common. In nematic phases, molecules align in a preferred direction defined by the director but lack positional order, allowing them to flow like a liquid while maintaining directional alignment. Due to their fluidity and sensitivity to external fields (such as electric or magnetic fields), nematic LCs are widely used in liquid crystal displays (LCDs). Smectic phases, in contrast, exhibit both orientational and partial positional order, with molecules organizing into layers that provide additional structural rigidity. Lyotropic liquid crystals, unlike thermotropic LCs, depend on solvent concentration and temperature for phase formation, with nematic order appearing in the simplest case \cite{deGennes1993,Collings2019}.

Nematic LCs are particularly significant due to their fundamental role in liquid crystal physics. When mixed with isotropic fluids or flexible polymers, thermotropic nematics exhibit phase behavior that depends on temperature and concentration, similar to lyotropic LCs. The phase behavior of nematic-isotropic fluid mixtures results from the interplay between molecular interactions, which promote phase separation, and entropic contributions, which favor mixing. Additionally, the presence of nematic-isotropic interfaces introduces unique phenomena, including anchoring effects, defect dynamics, and interfacial elasticity, all of which influence the kinetics of phase separation.

The phase behavior of binary isotropic mixtures is well understood, and classification schemes for phase diagrams have been established \cite{Konynenburg1980}. However, when one component is a nematic LC, these classifications become less applicable. The simplest scenario for nematic-isotropic mixtures involves a shift in the isotropic-nematic transition temperature with the nematic component concentration. More complex phase behavior arises when isotropic liquid-liquid phase separation also occurs. In this case, the temperature-concentration phase diagram resembles the solid-liquid-vapor phase diagram of a single-component system, with the nematic phase replacing the crystalline solid \cite{TelodaGama1984,Kato1996,Matsuyama2002}. Such behavior has been observed in mixtures of isotropic liquids \cite{Lagerwall2019,Guldin2018} or flexible polymers \cite{Orendi1989} with thermotropic LCs.

This review provides an overview of research on the phase separation and interfacial properties of nematic-isotropic fluid mixtures. It focuses on theoretical descriptions of the thermodynamic and kinetic behaviors that result from the coupling between concentration fluctuations and orientational order. The analysis is based on the simplest theoretical framework for describing phase separation in these systems: the combination of the Landau--de Gennes model \cite{deGennes1971,deGennes1993} for nematic ordering with the Cahn--Hilliard model \cite{Cahn1958,Cahn1961} for phase separation.

The Cahn--Hilliard model describes the contributions to free energy from fluid mixing, predicting phase diagrams based on material parameters and providing insight into miscibility and isotropic phase behavior. The Landau--de Gennes model accounts for the free energy contributions from nematic ordering, predicting the coexistence of nematic and isotropic phases and the nature of the phase transition.

The dynamics of phase separation or spinodal decomposition \cite{Cahn1961} in nematic-isotropic mixtures are governed by coupled conserved (concentration) and non-conserved (orientational ordering) fields and can be either isotropic or anisotropic \cite{Shen1995,Lin1997,Glotzer1999,Cates2002}. In the isotropic case, fluid phase separation occurs before orientational ordering, forming an initially isotropic phase-separated pattern. In the anisotropic case, ordered domains emerge as phase separation progresses. The kinetic pathway in the two-dimensional order parameter space is crucial for structure formation, even in the early stages \cite{Glotzer1999,Cates2002}. Furthermore, hydrodynamic effects can break the symmetry of the two phases through flow-alignment coupling, resulting in anisotropic domain growth away from the nematic ordering transition \cite{Araki2004}.

Advanced density functional theories \cite{Somoza1989,Osipov1993,Belli2014,Mederos2014} and molecular simulations \cite{Allen2019,Wilson2005,Zannoni2022} have expanded the scope of theoretical approaches by capturing the spatial and orientational correlations between nematic molecules. These studies have revealed the effects of molecular anisotropy on phase separation. Simulations, in particular, have highlighted the emergence of mesoscopic structures such as tactoids, which are a hallmark feature of nematic-isotropic systems \cite{Wang2018,vanderSchoot2022}.

Recently, research has shifted from passive to active nematic mixtures, where phase separation exhibits distinct characteristics due to the interplay between active forces, nematic order, and hydrodynamic interactions \cite{Yeomans2014,Yeomans2023,Marchetti2022,Marchetti2024,Dogic2022,Dogic2024,Coelho2023}. Unlike passive nematics, which follow equilibrium phase behavior, active nematics display non-equilibrium phenomena such as spontaneous flow, topological defect dynamics, and turbulence-like behavior. In active nematic mixtures, phase separation can be driven by activity-induced interfacial instabilities that affect domain growth and coarsening. Additionally, the presence of dynamic defects significantly influences the morphology and phase separation dynamics of the system. Although a comprehensive review of active nematic mixtures is beyond the scope of this work, a brief discussion is included to highlight the key differences between phase separation in passive and active nematic mixtures.

\section{Free Energy Functional}
The Landau free energy provides a powerful mean-field approach to describe phase transitions, using symmetry principles and power series expansions in an order parameter \cite{Landau1980}. It is widely used in condensed matter physics, material science, and statistical mechanics \cite{Toledano1987}.
Under the assumption of incompressibility, the two order parameters that characterize the phase transitions in mixtures of nematic and isotropic fluids are (proportional to) the concentration field, \( \phi(\mathbf{r}) \), and the nematic order parameter field, represented by the traceless, symmetric nematic tensor \( \mathbf{Q}(\mathbf{r}) \). 

The total free energy, \( \mathcal{F} \), of the nematic--isotropic fluid mixture may be written as the sum of a contribution from the concentration fluctuations, with free energy density \( f_\phi \), and a contribution from the fluctuations of the nematic order parameter, with free energy density \( f_Q \), and takes the form
\begin{equation}
    \mathcal{F} = \int_V \left[ f_\phi(\phi) + f_Q(\mathbf{Q}) + f_{\text{coupling}}(\phi, \mathbf{Q}) \right] \, dV,
    \label{FEFunctional}
\end{equation}
where  \( f_{\text{coupling}} \) models the coupling between the concentration and the nematic order parameter fields. 

Clearly, this free energy leaves out the description of other LC phase transitions, such as isotropic-smectic or nematic-semectic \cite{deGennes1993} but it is sufficient to describe the nematic--isotropic and the isotropic liquid--liquid transitions that occur in mixtures of a thermotropic liquid crystal with an isotropic fluid or flexible polymer \cite{Lagerwall2019,Guldin2018,Orendi1989,Shen1995}.

\subsection{Cahn--Hilliard Free Energy for Concentration}

The Cahn-Hilliard free energy describes phase separation and coarsening phenomena of systems with a scalar order parameter \cite{Cahn1958}. It describes how the free energy of an incompressible binary mixture depends on the concentration and its spatial variations.

The free energy of nematic--isotropic fluid mixtures arising from the concentration fluctuations is given by the Cahn--Hilliard free energy density, as the sum of bulk and gradient terms:
\begin{equation}
    f_\phi(\phi) = \frac{A}{2}\phi^2 + \frac{B}{4}\phi^4 + \frac{\kappa}{2} \left|\nabla \phi\right|^2,
    \label{phiFE}
\end{equation}
with the phenomenological material dependent coefficients, \( A < 0 \) in the two-phase region, \( B > 0 \) for stability, and  \( \kappa > 0 \) that penalizes sharp interfaces. 
The bulk free energy ensures that the system prefers certain values of the concentration, $\phi$, leading to phase separation below the critical point given by \( A = 0 \). The coefficient $A$ is generally written as $A=A_0(T-T_c)$ where $T-T_c$ stands for the concentration ordering field, usually the temperature measured with respect to the critical temperature, $T_c$. $A_0$  is a positive constant related to the energy scale of the concentration field. The coexisting order parameters are measured with respect to the critical concentration, and are given by $\phi_0= \pm \sqrt{\frac{-A}{B}}$, which are symmetrical and vanish at the critical point. The gradient energy term leads to the formation of diffuse interfaces and the total free energy favors configurations where the phases separate while minimizing the interfacial energy. The material dependent parameters may be set using experimental data but for illustration purposes we will set $A_0=B=1$ and measure $T_c$ in terms of the nematic-isotropic temperature of the pure nematic.

More realistic models, based on the Flory-Huggins \cite{Huggins1941,Flory1942,Doi2013} free energy density have been used in this context \cite{Matsuyama2002,Lagerwall2019,Shen1995,Liu1993} but Eq.~\eqref{phiFE} is particularly useful when considering non-uniform systems and interfacial phenomena.

\subsection{Landau--de Gennes Free Energy for Nematic Order}

For uniaxial nematics, the order is described by the director field $\mathbf n$, with cartesian components $n_i$, which gives the average direction of alignment of the nematic particles, and the scalar order parameter $S$, which measures the degree of alignment. These two fields are combined in a traceless symmetric tensor, which is written, $Q_{i j} = S(n_i n_j - \delta_{i j}/3)$ in 3 spatial dimensions \cite{deGennes1971}. Most bulk nematics are uniaxial but they may exhibit biaxiality under non-uniform conditions, at free interfaces \cite{Thurtell1985} or in the core of topological defects \cite{Kleman2006,Tasinkevych2012}.

The Landau-de Gennes (LdG) free energy density is given by an expansion in terms of the invariants of $\mathbf{Q}$, 
\begin{equation}
    f_Q(\mathbf{Q}) = \frac{a}{2}\,\Tr(\mathbf{Q}^2) - \frac{b}{3}\,\Tr(\mathbf{Q}^3) + \frac{c}{4}\,\left[\Tr(\mathbf{Q}^2)\right]^2 + \frac{L}{2} \left(\partial_k Q_{ij}\right)^2,
    \label{QFE}
    \end{equation}
with material dependent parameters \( a \), \( b \), and \( c \) (with \( a \) often taken as \( a = a_0 (T-T^*) \)). The coefficient of the elastic contribution \( L \) is related to the Frank elastic constants \cite{deGennes1993}. For simplicity, we consider a single-elastic constant, but discuss the effects of elastic anisotropy in later sections. 
The positive constant $a_0$ sets the energy scale of the nematic field and the positive elastic constant $L$ penalizes distortions of $Q_{ij}$. The coefficient $b$ is positive ensuring that the free energy exhibits a first-order phase transition, as observed in most nematics. 
 
The temperature $T^*$ marks the point where the isotropic phase loses its local stability, i.e., for $T < T^*$, the isotropic phase is unstable, meaning that small perturbations in the nematic order parameter will grow spontaneously, while for $T > T^*$, the isotropic phase remains metastable or even globally stable, depending on the relative free energy of the nematic and isotropic phases. Thus, $T^*$ is the lower isotropic spinodal temperature.

The nematic--isotropic (NI) transition temperature, denoted by $T_{NI}$, is determined by the condition that the free energies of the nematic and isotropic phases are equal:
\begin{equation}
 \mathcal{F}_{\text{iso}}(T_{NI}) = \mathcal{F}_{\text{nem}}(T_{NI}).
 \label{NIcoex}
\end{equation}
Since the transition is first-order, the isotropic phase remains metastable below $T_{NI}$, and the nematic phase remains metastable above $T_{NI}$. The upper nematic spinodal occurs at the temperature $T_{\text{up} }= T^* + \frac{b^2}{4ca_0} > T_{NI}$, where the nematic phase loses its metastability and transforms back into the isotropic phase.  

For uniaxial nematics in 3 dimensions, and assuming that spatial variations occur only in \(S\) (uniform \(\mathbf{n}\)) the free energy density simplifies to a similar function of $S$, with coefficents $a_S= \frac{2}{3} a$, $b_S=\frac{2}{9} b$, $c_S=\frac{2}{9} c$ and $L_S=\frac{2}{3} L$. For simplicity we will drop the subscript $S$ unless otherwise stated.

Again, more realistic models based on the Maier-Saupe or Onsager \cite{Maier1959,Maier1960,Onsager1949,Doi2013} free energy densities have been used \cite{Matsuyama2002} but Eq.~\eqref{QFE} is particularly useful when considering non-uniform systems.

\subsection{Coupling Between Concentration and Nematic Order}

The coupling term between the concentration field \( \phi \) and the nematic order parameter \( \mathbf{Q} \) is often modeled as:
\begin{equation}
    f_{\text{coupling}}(\phi, \mathbf{Q}) = - \lambda \, \phi \, \Tr(\mathbf{Q}^2),
    \label{couplingTerm}
\end{equation}
with coupling constant \( \lambda \). A positive \( \lambda \) indicates that regions with higher \(\phi\) favor stronger nematic ordering.

The full free energy density is then:
\begin{multline}
    f(\phi, \mathbf{Q}) = \frac{A}{2}\phi^2 + \frac{B}{4}\phi^4 + \frac{\kappa}{2} \left|\nabla \phi\right|^2 \\
    + \frac{a}{2}\,\Tr(\mathbf{Q}^2) - \frac{b}{3}\,\Tr(\mathbf{Q}^3) + \frac{c}{4}\,\left[\Tr(\mathbf{Q}^2)\right]^2 + \frac{L}{2}\left(\nabla \mathbf{Q}\right)^2 - \lambda \phi \,\Tr(\mathbf{Q}^2).
    \label{totalFE}
\end{multline}
Other models for the coupling may be and have been used. Popular choices consider a nematic free energy where the coefficients of Eq.~\eqref{QFE} become functions of $\phi$ \cite{Liu1993,Matsuyama2000,Yeomans2006,Morozov2023}. In these models the nematic ordering field is a function of the concentration, as expected for lyotropics. This accounts for the increase of nematic order with concentration of the nematogen, which is taken to increase with $\phi$. If the coefficients of the Cahn-Hilliard free energy, Eq.~\eqref{phiFE}, are constant \cite{Yeomans2006,Morozov2023} then the mixture exhibits a single transition from an isotropic liquid at low concentration to a nematic at high concentration.  

The model considered here has two thermodynamic ordering fields, one for concentration ($T$) and the other for nematic ordering (a linear combination of $T$ and $\phi$) and thus it can describe isotropic liquid-liquid phase separation followed by nematic ordering, at fixed temperature \cite{Shen1995,Matsuyama2000} as reported for mixtures of a thermotropic liquid crystal with an isotropic fluid \cite{Lagerwall2019,Guldin2018}.

\section{Analysis of the Phase Separation}

At the mean-field level, the equilibrium phases are determined by minimization of the free energy density, with respect to \( \phi \) and \( \mathbf{Q} \). In the isotropic phase (\( \mathbf{Q} = \mathbf{0} \)):
\begin{equation}
    f_{\text{iso}}(\phi) = \frac{A}{2}\phi^2 + \frac{B}{4}\phi^4.
    \label{bulkconcFE}
\end{equation}
The contribution to the free energy from orientational ordering of uniaxial nematics is:
\begin{equation}
    f_Q(\phi,S_0) = \frac{a}{2} S_0^2 - \frac{b}{3} S_0^3 + \frac{c}{4} S_0^4 - \lambda \phi S_0^2,
    \label{bulkSFE}
\end{equation}
where the (non-zero) equilibrium orientational order parameter, \( S_0 \), is found by solving:
\begin{equation}
    \frac{\partial f_Q}{\partial S} = a - b\,S_0 + c\,S_0^2 - 2\lambda\,\phi = 0.
    \label{S0}
\end{equation}
The total free energy density, $f(\phi)= f_{\text{iso}}(\phi) +f_Q(\phi,S_0)$, may be used to calculate the phase coexistence.

\subsection{Binodals}

The NI transition of the pure nematic is determined by Eq.~\eqref{NIcoex}, the solution of which gives $a_{NI} = \frac{2b^2}{9c}$. Substituting this in Eq.~\eqref{S0}, yields the value of the nematic scalar order parameter at the transition, $S_{NI} = \frac{2b}{3c}$. Adding an isotropic component will lower the transition temperature through the decrease of the nematic interactions. The coupling term, however, can stabilize the nematic order even at low concentrations. The shift of the transition temperature depends on both of these effects.

The phase diagram of the binary mixture is obtained by requiring chemical and mechanical equilibrium of the coexisting phases. The equilibrium concentrations in the isotropic and nematic phases, $\phi_I$ and $\phi_N$, are the solutions of the equations
for the balance of the chemical potential and the osmotic pressure, at a given temperature.

Balance of the chemical potential, $\mu = \frac{\delta f}{\delta \phi} = A\phi + B\phi^3 - \lambda S_0^2$, in the isotropic and nematic phases, requires: 
\begin{equation}
    \mu_I  = A\phi_I + B\phi_I^3 = \mu_N  = A\phi_N + B\phi_N^3 - \lambda S_0^2.
    \label{muImuN}
\end{equation}
Similarly, balance of the osmotic pressure, $\Pi= -f(\phi,S_0) + \phi \mu(\phi,S_0)$, in the coexisting phases, requires:
\begin{equation}
 \Pi_I= \frac{A}{2}\phi_I^2 + \frac{3B}{4}\phi_I^4 = \Pi_N=\frac{A}{2}\phi_N^2 + \frac{3B}{4}\phi_N^4 -\frac{a}{2} S_0^2 + \frac{b}{3} S_0^3 - \frac{c}{4} S_0^4,
 \label{PiIPiN}
\end{equation}
where the equilibrium order parameter, $S_0$, is the solution of Eq. \eqref{S0}.
The equilibrium conditions \eqref{muImuN} and \eqref{PiIPiN} are equivalent to a double tangent construction to the total free energy density, $f(\phi,S_0)$ \cite{Doi2013}.

\begin{figure}
  \center
  \includegraphics[width=1.0\linewidth]{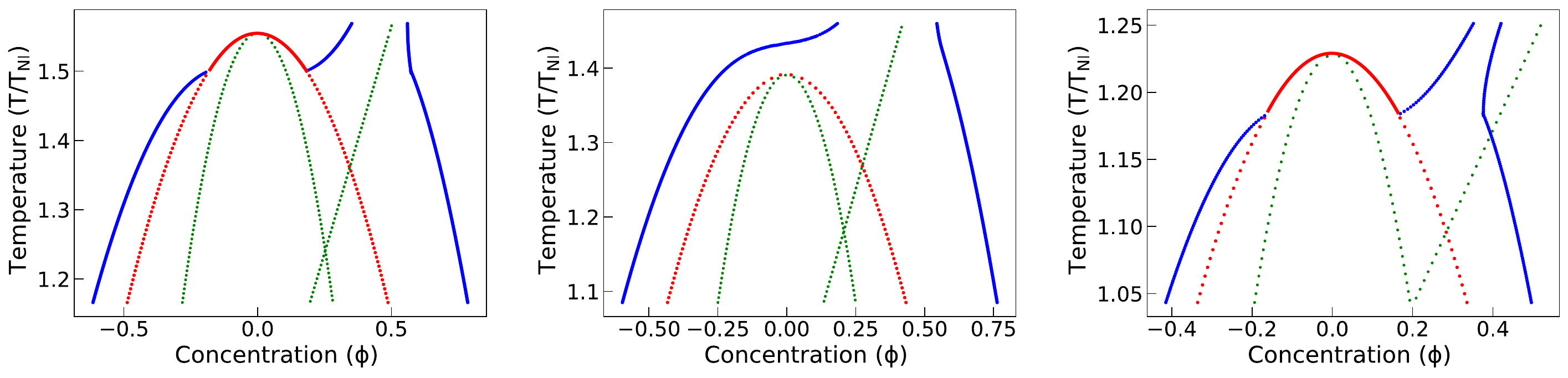} 
  \caption{Phase diagrams for the parameters: $a_0=1$, $b=1/2$, $c=1$, $A_0=1$, $B=1$, $T^*/T_{NI}=0.9$. (a) Triple point for a system with $T_c/T_{NI}=0.95$ and $\lambda=0.4$. (b) Single isotropic-nematic transition for $T_c/T_{NI}=0.85$ with the same $\lambda=0.4$. (c) Triple point for a lower coupling constant $\lambda=0.2$ and $T_c/T_{NI}=0.75$. Decreasing $\lambda$ destabilizes the nematic phase, while decreasing $T_c$ destabilizes the isotropic phase separation. The solid lines represent stable phase coexistence while the dotted red lines represent unstable or metastable isotropic coexistence. The green dotted lines are the isotropic spinodals that mark the instability of the isotropic phase to concentration (parabola) or nematic (straight line) fluctuations.}
  \label{fig:phase}
\end{figure}

The fluid phase diagram depends crucially on the ratio $T_c/T_{NI}$ and $\lambda$. The binodals of typical phase diagrams, and the isotropic spinodals, are depicted in Fig. \ref{fig:phase} for three sets of material parameters. The coefficients $a_0$, $A_0$ etc. are set to $1$, except $b$ that is set to $1/2$ to render the NI transition weakly first-order. In these units, the NI temperature of the pure nematic occurs at $T_{NI}=0.611$. The lower spinodal temperature was set to $T^*=0.9T_{NI}$. With this choice of parameters the phase diagrams depend on the ratio $T_c/T_{NI}$ and the value of the coupling constant, $\lambda$. Panel (a) exhibits a phase diagram with a triple point, where two isotropic phases coexist with a nematic phase. Above the triple point, isotropic phase separation occurs (solid red) before the isotropic-nematic transition (solid blue). Below the triple point there is a single isotropic-nematic transition (solid blue). Decreasing $T_c/T_{NI}$, for the same $\lambda$, renders the isotropic phase separation unstable and a single isotropic-nematic transition remains (panel (b) solid blue). In panel (c) we illustrate the effect of lowering the coupling constant $\lambda$. 

The phase diagram and the spinodal lines determine the phase separation dynamics for quenches from the isotropic phase, as we will discuss next. 

\subsection{Spinodals}
We now address the time evolution of the order parameters, $\phi$ and $\mathbf Q$, following a quench into the phase separated region. Since the concentration order parameter, $\phi$, is conserved, it follows a diffusion equation, while the orientational order parameter, $\mathbf Q$, follows non-conserved relaxational dynamics \cite{Cahn1958,Bray1994,Hohenberg1977,Ginzburg1950}:
\begin{equation}
\frac{\partial \phi}{\partial t} = \nabla \cdot \left( M_\phi \nabla \frac{\delta \mathcal F}{\delta \phi} \right), \quad \frac{\partial Q_{ij}}{\partial t} = -M_Q \frac{\delta \mathcal F }{\delta Q_{ij}}.
\label{CHLG}
\end{equation} 
where $M_\phi$ is the mobility of the concentration field and $M_Q$ is the relaxation coefficient of the orientational order field.

In the general case, for a biaxial $\mathbf Q$, there are 5 independent equations for $Q_{ij}$, which may be reduced to 3 in the uniaxial case. The last 3 can be written in terms of the scalar order parameter $S$ and the nematic director $\mathbf n$. In order to make analytical progress, we consider variations of $S$ only (neglect variations of $\mathbf n$), which is reasonable at nematic-isotropic interfaces. Then, the evolution equations are reduced to two scalar equations for $\phi$ and $S$:
\begin{equation}
\frac{\partial \phi}{\partial t} = M_\phi \nabla^2 \left(A\phi + B\phi^3 - \lambda S^2  - \kappa \nabla^2 \phi \right).
\label{phievol}
\end{equation}
\begin{equation}
\frac{\partial S}{\partial t} = -M_Q \left( a S - b S^2 + c S^3 - 2 \lambda \phi S - L \nabla^2 S \right).
\label{Sevol}
\end{equation}
We now expand the equations around a uniform equilibrium state $(\phi, S_0)$, assuming small perturbations of the form:
\begin{equation}
\phi(\mathbf{r}, t) = \phi + \delta \phi(\mathbf{r}, t) = \sum_{\mathbf{k}} \delta \phi_\mathbf{k} e^{i \mathbf{k} \cdot \mathbf{r} + \omega t}, \quad S (\mathbf{r}, t) = S_0 + \delta S(\mathbf{r}, t) =\sum_{\mathbf{k}} \delta S_\mathbf{k} e^{i \mathbf{k} \cdot \mathbf{r} + \omega t}
\label{perturbations}
\end{equation}
where $\omega$ is the growth rate of a perturbation with wavevector $\mathbf k$.
The linearized equations in Fourier space, and in matrix form are \cite{Bray1994,Langer1980,Langer1992,Cross1993,Chaikin2000}:
\begin{equation}
\begin{bmatrix}
-\omega - M_\phi k^2 (A + 3B\phi^2 + \kappa k^2) & 2 M_\phi k^2 \lambda S_0 \\
2 M_Q \lambda S_0 & -\omega - M_Q (a - 2 \lambda \phi - 2 bS_0+ 3c S_0^2 -L k^2)
\end{bmatrix}
\begin{bmatrix}
\delta \phi_k \\
\delta S_k
\end{bmatrix}
=
\begin{bmatrix}
0 \\
0
\end{bmatrix}.
\label{matrixevolution}
\end{equation}

The dispersion relation for the growth rate of fluctuations, $\omega(\mathbf k)$, is obtained by setting the determinant of the matrix of coefficients of the linearized equations to zero.
The spinodal is given by the onset of instability, where the largest growth rate vanishes, 
\begin{equation}
(A + 3 B \phi^2 + \kappa k^2)(a - 2 \lambda \phi - 2 b S_0 + 3 c S_0^2 - L k^2) = 4 \lambda^2 S_0^2.
\label{spinodal}
\end{equation}
The solution of Eq.~\eqref{spinodal} gives the critical wavenumber $k_c$ for which the system becomes unstable. If a real, positive $k$ satisfies Eq.~\eqref{spinodal} the system enters the spinodal region with the emergence of pattern formation. More generally, the spinodal defines the boundary where the homogeneous state becomes linearly unstable, i.e., where at least one eigenvalue of the stability matrix becomes zero. If this occurs at $k=0$ the system undergoes bulk (isotropic or nematic) phase separation. Note that the spinodal equation for $k=0$, is the stability condition for the bulk free energy, where the determinant of the Hessian vanishes. 
The equation for the spinodal highligts the coupled nature of the instabilities and their dependence on the coupling term $\lambda$. There is an isotropic spinodal region where only concentration fluctuations grow, following classic Cahn-Hilliard dynamics (green dotted parabola in Fig. \ref{fig:phase}). In the anisotropic spinodal region (green dotted straight line in Fig. \ref{fig:phase}) concentration and orientation fluctuations grow, but at different rates, which depend on the mobility coefficients \cite{Lin1997,Liu1996}. There is also a nematic spinodal, which is the solution of Eq.~\eqref{spinodal} for non-zero $S_0$, close to the nematic binodal (not shown). 
In general, the evolution of $\mathbf Q$ given by Eq.~\eqref{CHLG} has to be considered \cite{TenBosch1991}, to describe the dependence of the phase separation dynamics on the anisotropy of the elastic free energy and the interfacial anchoring. Studies revealed \cite{Glotzer1999,Ball1991,Cates2000} that spinodal decomposition may be significantly influenced by the presence of orientational order, even in the early stage \cite{Cates2000}. Elastic distortions introduce anisotropic domain shapes, leading to preferred growth directions and to the emergence of fribillar networks of the minority phase \cite{Glotzer1999,Ball1991}. If the director field is aligned along a specific axis, phase separation occurs preferentially in that direction, elongating the ordered domains \cite{Araki2006}. Hydrodynamic effects also affect the phase separation dynamics through flow alignment coupling \cite{Araki2004}. Detailed hydrodynamic studies of passive nematic and isotropic mixtures are scarce \cite{Araki2004,Yeomans2006,Morozov2023} when compared to the hydrodynamic studies of active nematic mixtures (to be addressed later), where spontaneous flows occur even under steady state conditions \cite{Yeomans2014,Yeomans2023,Dogic2022,Dogic2024,Coelho2023,Marchetti2013,Cates2018,Sagues2018a}. In the next section, we outline the hydrodynamic equations and their coupling to the thermodynamics of passive mixtures. 

\section{Hydrodynamic coupling}

Flowing nematics are often described by continuum hydrodynamic theories. A widely used model was proposed by Beris and Edwards~\cite{beris1994thermodynamics}. This model extends classical fluid dynamics by incorporating the effects of orientational order and elasticity. The model considers that the orientation of the nematic described by the $\mathbf Q$ tensor is coupled to the fluid velocity, \( \mathbf{v}(\mathbf{r},t) \). The key equations governing the dynamics are the continuity and Navier-Stokes equations, which describe the time evolution of the velocity field, and the Beris-Edwards equation that describes the evolution of $\mathbf Q$.

The continuity and Navier-Stokes equations for an incompressible fluid are given by:
\begin{eqnarray}
    \nabla \cdot \mathbf{v} = 0, \quad \text{and} \quad     \rho \frac{D \mathbf{v}}{Dt} = \nabla \cdot \bm{\Pi},
    \label{NS-cont-eq}
\end{eqnarray}
where $D/Dt=\partial/\partial t + \mathbf{v}\cdot \nabla$ is the material derivative and \( \bm{\Pi} \) is the total stress tensor, which includes contributions from elastic, viscous, and pressure terms.

The Beris-Edwards equation for the uniaxial order parameter \( \mathbf{Q} \) reads:
\begin{equation}
    \frac{D \mathbf{Q}}{D t} - \mathbf{S} = \Gamma_Q \mathbf{H},
    \label{BE-eq}
\end{equation}
where \( \mathbf{S} \) is the co-rotational term, which accounts for flow alignment and tumbling effects, \( \Gamma_Q \) is the rotational viscosity and \( \mathbf{H} \) is the molecular field, which describes the relaxation of $\mathbf{Q}$ towards the minimum of the free energy:
\begin{equation}
H_{ij} = -\frac{\delta \mathcal{F} }{\delta Q_{ij}} + \frac{\delta_{ij}}{D} \Tr \left( \frac{\delta \mathcal{F}}{\delta Q_{kl}}  \right).
\end{equation}
The co-rotational term is:
\begin{equation}
    S_{ij} = ( \xi A_{ik} + W_{ik})\left( Q_{ij} + \frac{\delta_{ik}}{D}   \right) + \left( Q_{ik}+\frac{\delta_{ik}}{D} \right) \left( \xi A_{kj} - W_{kj} \right) - 2\xi \left( Q_{ij}+\frac{\delta_{ij}}{D} \right) Q_{kl} \partial_k v_l
\end{equation}
Here, $D$ is the spatial dimension and \( \xi \) is the alignment parameter, which determines how the $\mathrm Q$ tensor reacts to the symmetric and antisymmetric parts of the velocity gradient tensor, the shear rate $A_{ij}=(\partial_i v_j + \partial_j v_i)/2$, and the vorticity, $W_{ij}=(\partial_j v_i - \partial_i v_j)/2$. $\xi$ is determined by the shape of the particles, with $\xi>0$ for rod-like particles and $\xi<0$ for disk-like particles. The stress tensor in the Navier-Stokes equation is:
\begin{equation}
    \Pi_{ij} = -p \delta_{ij} + 2\eta A_{ij} + \sigma_{ij}^{\text{elastic}} ,
\end{equation}
where $p$ stands for the hydrostatic pressure. The elastic stress contribution is:
\begin{align}
\sigma_{ij}^{\text{elastic}} &=  2\xi \left( Q_{ij} +\frac{\delta_{ij}}{D} \right)Q_{kl}H_{kl}  
- \xi H_{ik} \left( Q_{kj}+\frac{\delta_{kj}}{D} \right) - \xi \left( Q_{ik} +\frac{\delta_{ik}}{D} \right) H_{k j}\nonumber \\ & 
- \frac{\delta \mathcal{F}}{\delta (\partial_j Q_{k\nu})}\, \partial _i Q_{k\nu}  + Q_{ik}H_{kj} - H_{ik}Q_{kj} .
\end{align}

This set of equations describes the dynamics of flowing nematics with non-uniform scalar order parameter, $S$. When $S$ is uniform (deep in the nematic), the simpler Ericksen-Leslie equations~\cite{Stewart2019} may be used. In mixtures of immiscible fluids, where one is isotropic and the other is nematic or isotropic, the Beris-Edwards equations are coupled to the Cahn-Hilliard equation, Eq.~\eqref{CHLG}, which becomes
\begin{equation}
\frac{D \phi}{D t} = \nabla \cdot \left( M_\phi \nabla \frac{\delta \mathcal F}{\delta \phi} \right).
\label{CH-hydro}
\end{equation} 
The hydrodynamic equations can be solved analytically under simplifying assumptions or numerically using a range of different methods. One common approach is to use the lattice Boltzmann method to solve Eq.~\eqref{NS-cont-eq} and \eqref{CH-hydro} and finite-differences to solve Eq.~\eqref{BE-eq}, although other combinations of methods may be found in the literature~\cite{Krger2017, denniston2004lattice, C6SM01275B, PhysRevE.74.041708}. Coupling the nematic relaxation with hydrodynamics is essential in problems with forced mass flows~\cite{PhysRevResearch.5.033210} and in studies of active nematics where spontaneous flows occur even in the steady state~\cite{Coelho2023,Doostmohammadi2018,Singh2024}.

\section{Interfacial Phenomena}

The interplay of surface alignment and bulk elastic forces gives rise to interfacial phenomena that influence the dynamics of phase separation. Molecular interactions, in particular steric forces, determine the strength and orientation of surface alignment \cite{deGennes1971,Thurtell1985,Jerome1991,MartindelRio1995}. Furthermore, interfacial curvature can lead to elastic distortions producing spontaneous patterns and instabilities in confined geometries \cite{deGennes1993,Chaikin2000,Musevic2017,Lavrentovich2020}. Studies highlighted the importance of interface roughness and fluctuations, which arise from thermal effects. In nematics, the coupling of thermal fluctuations to the orientational order suppresses short-wavelength capillary waves \cite{Schmid2005,Schmid2001}. In active systems, interfacial flows driven by activity enhance interfacial fluctuations and may lead to instabilities \cite{Marchetti2024,Dogic2022,Dogic2024}. 

\subsection{Surface Tension}

We start by considering a planar interface of a phase separated binary mixture described by the Cahn-Hilliard free energy, Eq. \eqref{phiFE} \cite{Cahn1958,Cahn1961}. The solution of the Euler-Lagrange equation yields the well known equilibrium hyperbolic tangent profile, $\phi(z)$:
\begin{equation}
    \kappa \frac{d^2 \phi}{dz^2} = A\phi + B\phi^3,  \quad  \phi(z) = \phi_0 \tanh\left(\frac{z}{\xi_\phi}\right),    
    \label{ELphi}
\end{equation}
where the coexisting concentrations are $\phi_0^2 = -\frac{A}{B}$. The characteristic interface width is given by $ \xi_\phi = \sqrt{\frac{2\kappa}{|A|}}$.
The surface tension is then obtained:
\begin{equation}
    \gamma_\phi = \int_{-\infty}^{\infty} \kappa \left( \frac{d\phi}{dz} \right)^2 dz =\frac{4}{3} \frac{ \kappa \phi_0^2}{\xi_\phi},   
    \label{gammaphi}
    \end{equation}
and is found to scale with $\sqrt{\kappa}$ and $|A|^{\frac{3}{2}}$. At the critical point, the surface tension vanishes as $\gamma_\phi \propto (T_c-T)^{\frac{3}{2}}$ \cite{Widombook}.

At the NI interface described by the Landau-de Gennes model \cite{deGennes1971}, the coexisting order parameters are $S=0$ and $S_0=\frac{2b}{3c}$, the coupling is $a_{NI}=\frac{2b^2}{9c}$ and we can write the free energy density, $f(S)=\frac{a}{2}S^2-\frac{b}{3}S^3+\frac{c}{4}S^4=\frac{a}{2}S^2\Bigl(1-\frac{b}{3a}S\Bigr)^2$. Then, the first integral of the Euler--Lagrange equation gives,
\begin{equation}
\frac{L}{2}\Biggl(\frac{dS}{dz}\Biggr)^2 = f(S)-f(S_0) = \frac{a}{2} S^2 \Biggl (1-\frac{b}{3a}S\Biggr)^2.
\label{dS/dzsq}
\end{equation}
and the equilibrium orientational order parameter profile, $S(z)$, is:
\begin{equation}
S(z)= \frac{1}{2}S_0 \Biggl (1+\mathrm{tanh} \Biggl (\frac{z}{\xi_S} \Biggl ) \Biggr )
\label{Sz}
\end{equation}
with the characteristic interface width, $\xi_S = 2 \sqrt{\frac {L}{a}}$. The surface tension is then obtained:
\begin{equation}
\gamma_S=L\int_{-\infty}^{\infty} \Biggl(\frac{dS}{dz} \Biggr )^2dz = \frac {L S_0^2}{3\xi_S},
\label{gammaS}
\end{equation}
and is found to scale with $\sqrt{L}$ and $|a|^{\frac{3}{2}}$. Since the NI transition is weakly first-order, $|a|$ is generally much smaller than $|A|$ and $\gamma_S \ll \gamma_\phi$.
The surface tension of nematic-isotropic mixtures is not amenable to analytical solution but 
may be calculated numerically and is generally dominated by $\gamma_\phi$ \cite{Cates2002,Coelho2023}.   

\subsection{Interfacial Anchoring}

In the bulk of a nematic liquid crystal, the orientation of the director field \( \mathbf{n} \) is arbitrary, as the free energy depends only on the degree of orientational order, not on the direction of alignment. However, the presence of boundaries breaks translational invariance, which in turn breaks the rotational symmetry of the nematic phase. This symmetry breaking manifests at free surfaces and is referred to as interfacial anchoring~\cite{deGennes1971, Jerome1991}.

Interfacial anchoring originates from the elastic anisotropy of the nematic phase—that is, the difference between the splay, twist, and bend elastic constants~\cite{deGennes1993}. In the Landau--de Gennes formalism, the elastic free energy density can be expressed in terms of three independent quadratic invariants with coefficients \( L_1 \), \( L_2 \), and \( L_3 \):
\begin{equation}
    f_{\mathrm{elastic}} = \frac{L_1}{2} \partial_k Q_{ij} \partial_k Q_{ij} + \frac{L_2}{2} \partial_j Q_{ij} \partial_k Q_{ik} + \frac{L_3}{2} \partial_j Q_{ik} \partial_k Q_{ij},
    \label{fel_full}
\end{equation}
where \( Q_{ij} \) is the nematic order parameter tensor. The second and third terms differ by a total derivative and contribute a surface term upon integration over the volume. Under suitable conditions, such boundary contributions can be neglected or treated separately. As a result, the effective bulk elastic energy depends on the combination \( L_2 + L_3 \) rather than on the individual coefficients.

At a free surface, the uniaxial elastic free energy density can be approximated as~\cite{deGennes1971}:
\begin{equation}
    f_{\mathrm{interface}} \sim \frac{S_0^2}{2} \left[ L_1 + L_2 \cos^2 \theta \right] \left( \frac{dS}{dz} \right)^2,
\end{equation}
where \( S_0 \) is the bulk scalar order parameter, \( \theta \) is the angle between the director \( \mathbf{n} \) and the surface normal (assumed to be along the \( z \)-axis), and \( L_2 \) and \( L_3 \) are combined into an effective \( L_2 \). Minimizing \( f_{\mathrm{interface}} \) with respect to \( \theta \) shows that the preferred anchoring depends on the sign of \( L_2 \):  For \( L_2 > 0 \), the minimum occurs at \( \theta = 0 \), corresponding to homeotropic anchoring (alignment perpendicular to the surface). For \( L_2 < 0 \), the minimum is at \( \theta = \pi/2 \), corresponding to planar anchoring (alignment parallel to the surface)~\cite{deGennes1971, Coelho2021ptrsa}.  
In the special case where \( L_2 = 0 \), the elastic energy is isotropic with respect to the director orientation, leading to degenerate anchoring.

When using a simplified one-constant approximation for the elastic free energy, interfacial anchoring is often modeled using an additional surface energy term. A commonly used form is:
\begin{equation}
    \mathcal{F}_{\text{anchoring}} = \frac{W}{2} \int_A \left[ 1 - \cos^2 \theta \right] \, dA,
\end{equation}
where \( W \) is the anchoring strength coefficient and the integral is over the interfacial area \( A \). A positive value of \( W \) favors homeotropic anchoring, while a negative \( W \) favors planar anchoring. In nematic--isotropic mixtures, interfacial anchoring arises similarly due to the coupling between the nematic director and the local interface normal, as in pure nematics.

\subsection{Nematic Defects and Emulsions}
\begin{figure}
  \center
  \includegraphics[width=\linewidth]{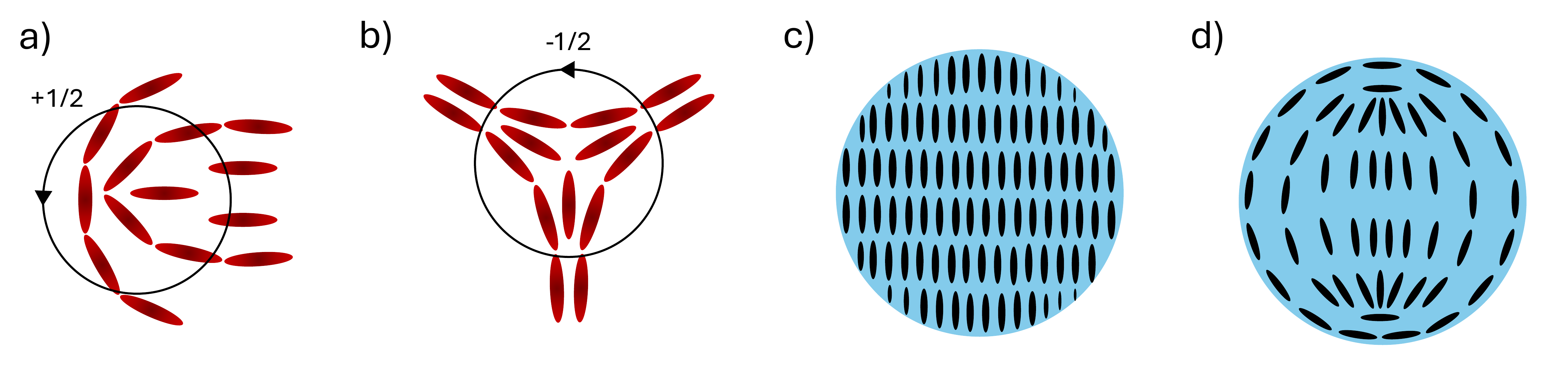} 
  \caption{Schematic representation of defects in 2D nematics. (a) $+1/2$ defect. (b) $-1/2$ defect. (c) Nematic droplet without distortions. (d) Nematic droplet with two $+1/2$ defects that will evolve \cite{Lavrentovich2023}.  }
  \label{fig:defects}
\end{figure}

The presence of orientational order may affect the shape of nematic droplets. The elastic free energy associated with director distortions inside the droplet competes with the interfacial free energy to determine the droplet equilibrium shape. If surface anchoring is weak, the droplet remains nearly spherical due to the dominance of the isotropic surface tension (see Fig. \ref{fig:defects} panel (c)). Under strong anchoring non-spherical shapes may arise as the nematic ordering resists the isotropic tendency of the surface tension. In addition, nematic droplets often exhibit topological defects due to the conflicting constraints of bulk elasticity and surface anchoring (see Fig.\ref{fig:defects} panels (a), (b) and (d))   \cite{Chaikin2000,Musevic2017,Kurik1988}. 

Topological defects in nematic liquid crystals are singularities in the director field where the local orientational order is disrupted. These defects are classified using homotopy theory~\cite{Kurik1988, Mermin1979}. The primary types include point defects (or hedgehogs), where the director field $\mathbf{n}$ radiates inward or outward, and line defects (or disclinations), where $\mathbf{n}$ undergoes a rotation along a line~\cite{Kurik1988, Mermin1979}.

Disclinations are characterized by a topological charge (or winding number), defined as:
\begin{equation}
    m = \frac{1}{2\pi} \oint_{\Gamma} d\theta,
\end{equation}
where $\theta$ is the angle of the director in a plane perpendicular to the defect line, and the integral is taken along a closed loop $\Gamma$ encircling the defect. The most common and stable disclinations have charges $m = \pm \tfrac{1}{2}$ (see Fig.~\ref{fig:defects}, panels (a) and (b)), while higher-order disclinations with $m = \pm 1$ are typically unstable and tend to split into pairs of half-integer defects~\cite{Kurik1988, Mermin1979}.

Point defects are similarly characterized by a topological charge~\cite{Mermin1979, Kurik1988}:
\begin{equation}
    q = \frac{1}{8\pi} \int_{\partial V} \epsilon_{ijk} n_i \frac{\partial n_j}{\partial x_k} \, dS,
\end{equation}
where $\epsilon_{ijk}$ is the Levi-Civita antisymmetric tensor and the surface integral is taken over a closed surface enclosing the defect. Common examples include radial hedgehogs, with $\mathbf{n} \sim \mathbf{r}/|\mathbf{r}|$ and charge $q = +1$, and hyperbolic hedgehogs, which exhibit a saddle-like structure and carry charge $q = -1$~\cite{Kurik1988, Mermin1979}.

The evolution of topological defects is governed by the Landau--de Gennes free energy and the nematic hydrodynamics described earlier. Defects with opposite topological charge attract and annihilate, while like-charged defects repel one another (see Fig.~\ref{fig:defects}). In passive nematics, defects relax to equilibrium configurations and remain static. In contrast, in active nematics, defects become motile and give rise to novel dynamical steady states~\cite{Giomi2012, Marchetti2013a, Marchetti2022a, Sagues2018b, Dogic2020, Duclos2024}.

Topological defects can be investigated in nematic emulsions—dispersions of nematic droplets suspended in a continuous isotropic liquid—where parameters such as droplet size, surface anchoring, and stability can be precisely controlled~\cite{Musevic2017}. A specific class of these systems, known as inverse nematic emulsions, features a continuous nematic phase surrounding dispersed isotropic droplets~\cite{Musevic2017, Weitz1999, Stark2001}. The interaction between the nematic field and the isotropic inclusions gives rise to complex defect textures, and the droplets experience elastic forces mediated by the surrounding nematic~\cite{Musevic2017, Stark2001}. These tunable, defect-mediated interactions can drive the self-assembly of ordered structures, offering promising applications across various fields~\cite{Musevic2017}.

\subsection{Interfacial Fluctuations}
Thermal fluctuations play a crucial role in shaping the structure of fluid interfaces and influence the properties of phase-separated fluids. A fluctuating interface between two phases can be described by a height field \( h(\mathbf{r}_\parallel) \), where \( \mathbf{r}_\parallel \) denotes coordinates in the plane of the average interface. This height function captures deviations from the mean interfacial position due to thermal undulations.

For small amplitude fluctuations, the effective interfacial energy can be approximated by a harmonic (quadratic) Hamiltonian~\cite{Chaikin2000, Stillinger1965, Safran1994}:
\begin{equation}
    \mathcal{H}_{\text{cw}} = \frac{\gamma}{2} \int d \mathbf{r}_\parallel\, \left( \nabla h(\mathbf{r}_\parallel) \right)^2 = \frac{\gamma}{2} \sum_{\mathbf{k}} k^2 \, |h(\mathbf{k})|^2,
\end{equation}
where \( \gamma \) is the interfacial tension, and the Fourier transform of the height field is given by
\[
h(\mathbf{r}_\parallel) = \sum_{\mathbf{k}} h(\mathbf{k}) e^{i \mathbf{k} \cdot \mathbf{r}_\parallel}.
\]

At equilibrium, the equipartition theorem dictates that each mode has an average energy of \( \frac{1}{2} k_B T \), which leads to the capillary wave fluctuation spectrum:
\begin{equation}
    \langle |h(\mathbf{k})|^2 \rangle = \frac{k_B T}{\gamma k^2},
\end{equation}
revealing that long wavelength fluctuations (small \( k \)) render the interface rough \cite{Stillinger1965}.

In nematic--isotropic mixtures, anchoring conditions at the interface and the elastic energy associated with director distortions modify the effective interfacial tension, thereby influencing the fluctuation spectrum~\cite{Schmid2005, Schmid2001}. This anisotropy leads to direction-dependent roughness at short wavelengths, which can affect interfacial phenomena such as droplet coalescence and defect dynamics. Moreover, the coupling between the orientational order parameter and the interfacial height field introduces an additional bending energy term in the interfacial Hamiltonian~\cite{Chaikin2000, Safran1994, Helfrich1973}.

The order parameter profile near a fluctuating interface can be approximated by a shifted version of the mean-field profile corresponding to a flat interface. Expanding this shifted profile to first order in the interfacial height fluctuation \( h(\mathbf{r}_\parallel) \), we obtain:
\begin{equation}
    Q_{ij}(x, y, z) \approx Q_{ij}^0(z - h(\mathbf{r}_\parallel)) \approx Q_{ij}^0(z) - h(\mathbf{r}_\parallel) \frac{dQ_{ij}^0}{dz},
\end{equation}
where \( Q_{ij}^0(z) \) denotes the equilibrium order parameter profile in the absence of fluctuations.

Taking the derivative with respect to \( z \), the gradient of the order parameter becomes:
\begin{equation}
    \frac{dQ_{ij}}{dz} \approx \frac{dQ_{ij}^0}{dz} - h(\mathbf{r}_\parallel) \frac{d^2 Q_{ij}^0}{dz^2}.
\end{equation}
This expression reveals that the interfacial height fluctuations couple linearly to the curvature (second derivative) of the equilibrium order parameter profile, thereby modifying the local structure of the interface.

The elastic energy associated with interfacial fluctuations can be expressed as:
\begin{equation}
    \mathcal{H}_{\text{elastic}} = \frac{L_{\text{eff}}}{2} \sum_{\mathbf{k}} k^4 \, |h(\mathbf{k})|^2,
\end{equation}
where \( L_{\text{eff}} \) is an effective elastic modulus that depends on the elastic constant \( L \) and the gradient of the equilibrium order parameter profile.

The fluctuation spectrum of the interface can be derived using the equipartition theorem~\cite{Chaikin2000, Safran1994, Helfrich1973}, yielding:
\begin{equation}
    \langle |h(\mathbf{k})|^2 \rangle = \frac{k_B T}{\gamma k^2 + L_{\text{eff}} k^4},
\end{equation}
where \( k_B T \) is the thermal energy, \( \gamma \) is the interfacial tension, and \( L_{\text{eff}} \) is the effective bending modulus introduced previously. 

At long wavelengths (small \( k \)), the interfacial tension dominates the spectrum, leading to \( \langle |h(\mathbf{k})|^2 \rangle \sim 1/\gamma k^2 \), whereas at short wavelengths (large \( k \)), bending rigidity becomes significant and suppresses fluctuations, resulting in \( \langle |h(\mathbf{k})|^2 \rangle \sim 1/L_{\text{eff}} k^4 \). Although the bending term suppresses short-wavelength fluctuations, the large-scale roughness of the interface remains governed by the interfacial tension. In nematic--isotropic mixtures, coupling between the nematic order and interfacial fluctuations is expected to lead to similar effects. 

\section{Active Nematic Mixtures}

\begin{figure}
  \center
  \includegraphics[width=\linewidth]{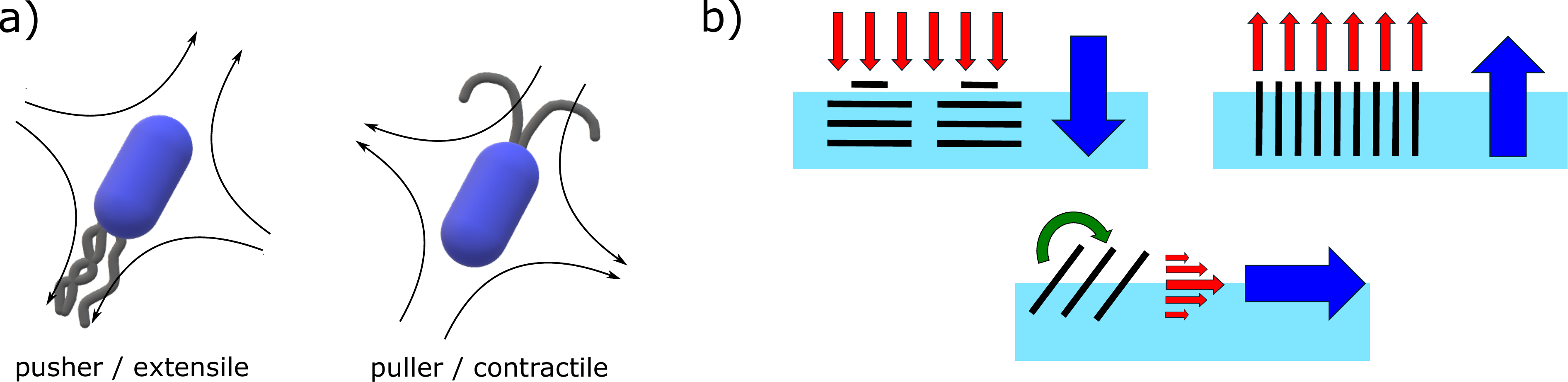} 
  \caption{Active nematics. (a) Representation of microswimmers: pushers (extensile systems) and pullers (contractile systems). (b) Normal and tangential components of the active force at a nematic-isotropic interface for extensile systems. The dark blue arrow represents the active force while the red ones represent the flow. Bars represent the nematic director.}
  \label{fig:active}
\end{figure}

Active nematics are dense systems of rod-like particles that exhibit nematic orientational order while being continuously driven out of equilibrium by energy input at the microscopic scale (see Fig. \ref{fig:active} panel (a))\cite{Ramaswamy2010,MarchettiRMP2013}. Unlike passive systems, which require external forces to induce flow, active nematics are internally driven by active stresses, leading to novel dynamical steady states. These include giant density fluctuations \cite{Ramaswamy2010,MarchettiRMP2013}, motile topological defects \cite{Giomi2012,Marchetti2013a,Sagues2018b,Sagues2020}, and spontaneous flow states \cite{Alert2022b,Thampi2022}, which are absent in passive nematics. Generally, long-range nematic order is disrupted by activity, making initially ordered active nematics unstable and driving them into globally disordered states characterized by chaotic flows, known as active turbulence \cite{Alert2022b}.

In active nematic mixtures, the motion of the nematic constituents disrupts equilibrium-ordered states, resulting in dynamic steady states where phase-separated domains continuously reorganize due to activity-driven hydrodynamic instabilities. Numerical simulations and experiments have demonstrated that activity influences phase separation boundaries and domain dynamics, affecting their size, morphology, and stability \cite{Marchetti2022,Dogic2022,Dogic2024,Doostmohammadi2024}.

\subsection{Incorporating Activity}

To model active nematic mixtures, additional terms are introduced in the hydrodynamic equations to account for active stresses. A common approach modifies the stress tensor by incorporating an active stress term \cite{Ramaswamy2002,Ramaswamy2003}:
\begin{equation}
    \Pi_{ij}^{\text{active}} = - \zeta \, Q_{ij},
    \label{activestress}
\end{equation}
where \( \zeta \) is the activity parameter. The sign of \( \zeta \) distinguishes between extensile (\( \zeta > 0 \)) and contractile (\( \zeta < 0 \)) systems. In extensile systems, swimmers push fluid outward along their swimming axis and pull it inward from the sides, creating a dipolar flow field with outward thrust at the front and rear (see Fig. \ref{fig:active} panel (a)). Examples include bacteria (e.g., \textit{E. coli}), cytoskeletal networks with motor proteins, and self-propelled rods. In contractile systems, swimmers pull fluid inward along the swimming axis and push it outward from the sides, generating a dipolar flow field with inward forces at the front and rear (see Fig. \ref{fig:active} panel (a)). Examples include algae (e.g., \textit{Chlamydomonas}) and contractile actomyosin networks.

Activity modifies the hydrodynamic equations, through the pressure tensor \( \bm{\Pi} \) in the Navier-Stokes equation \eqref{NS-cont-eq}, which includes the additional term \( \bm{\Pi}^{\text{active}} \). The active stress enters the momentum equation as a continuously applied stress. Even in the absence of external forces, a sufficiently large \( |\zeta| \) can destabilize the quiescent state, as originally predicted by Simha and Ramaswamy’s linear stability analysis \cite{Ramaswamy2002}. A uniformly aligned active nematic is unstable beyond a critical activity threshold, which can approach zero in large systems, leading to spontaneous flow. This instability arises because active stresses overcome the passive elastic and viscous restoring torques, triggering long-wavelength fluid motion and leading to active turbulence \cite{Alert2022b}.

\subsection{Phase Separation and Interfacial Phenomena}

\begin{figure}
  \center
  \includegraphics[width=0.6\linewidth]{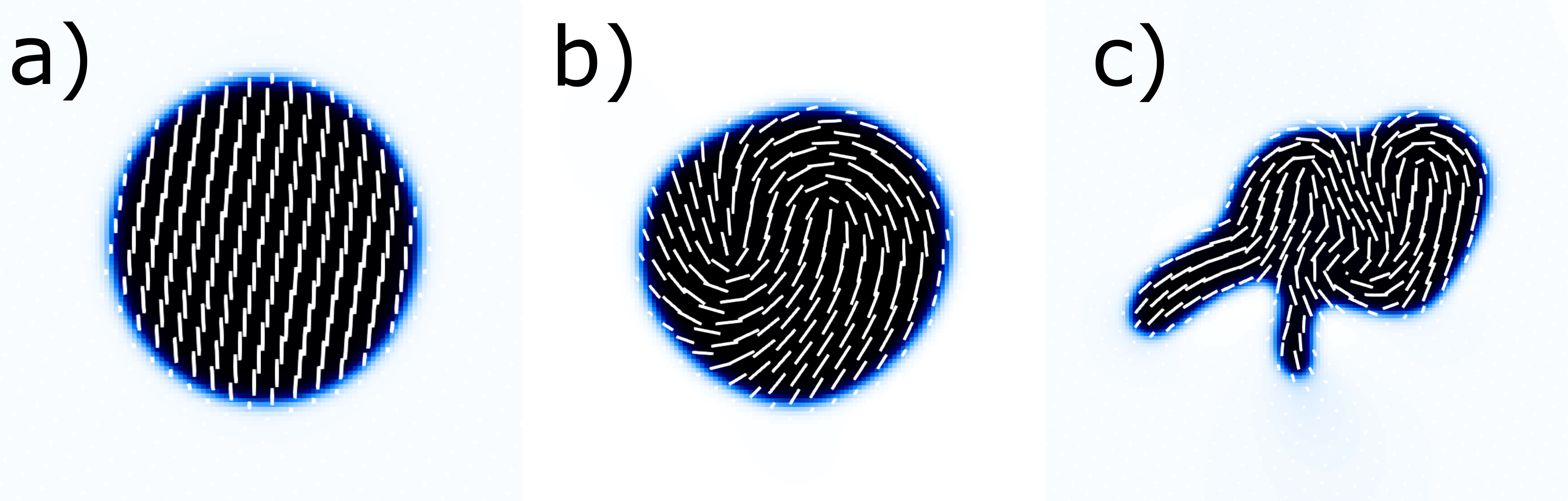} 
  \caption{Extensile active nematic droplets. (a) Stable droplet at low activity $\zeta=0.001$. (b) Oscillating droplet with periodic deformations at intermediate activity, $\zeta=0.01$. (c) Chaotic droplet with chaotic deformations at high activity, $\zeta=0.05$. Black depicts the nematic with the director field represented by white bars, while white depicts the isotropic fluid. The model and parameters are the same as those used in Ref.\cite{Coelho2023}.}
  \label{fig:droplets}
\end{figure}

In passive systems, the nematic-isotropic transition is governed by the balance of enthalpic and entropic effects. In active nematics, the interplay between elastic interactions and activity determines whether the system remains isotropic, orders nematically, or transitions into the active turbulent state. Active stresses coupled with fluid dynamics can enhance nematic order via flow alignment \cite{ThampiEPL2015,Thampi2020} or destabilize it by generating active turbulence \cite{Alert2022b}. Phase separation in nematic-isotropic mixtures is also influenced by activity, which alters phase boundaries and steady-state dynamics \cite{Marchetti2022,Dogic2023}.

\begin{figure}
  \center
  \includegraphics[width=0.6\linewidth]{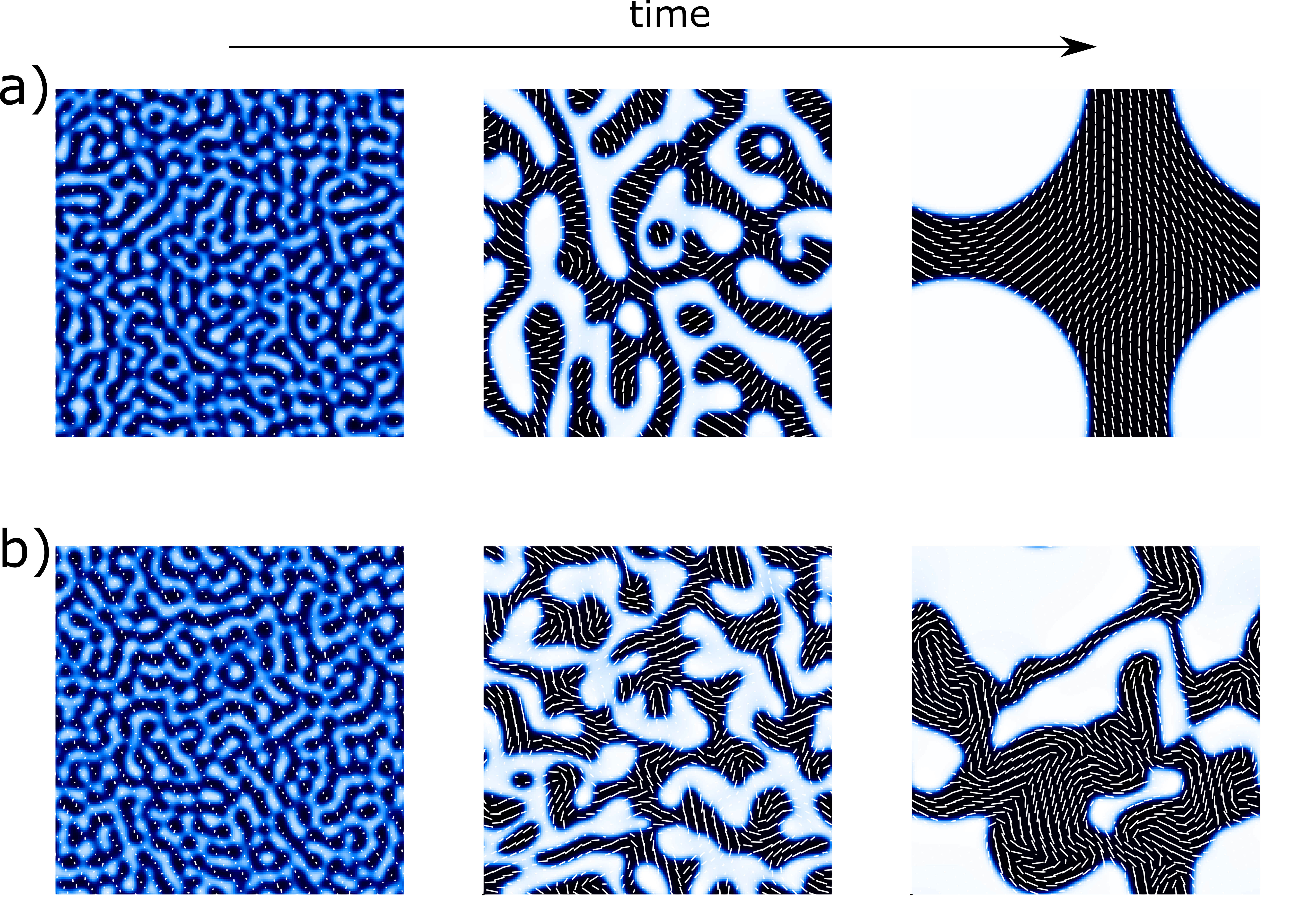} 
  \caption{Spinodal decomposition of an active emulsion at (a) low activity, $\zeta=0.001$ and (b) high activity $\zeta=0.05$. Black depicts the nematic with the director represented by white bars while white depicts the isotropic fluid. The model and parameters are the same as those in Ref.\cite{Coelho2023}. }
  \label{fig:emulsion}
\end{figure}

Activity can produce nontrivial interfacial structures, such as traveling waves or oscillatory patterns \cite{Marchetti2024,Dogic2022}. Unlike classical passive mixtures, where surface tension stabilizes interfaces, active mixtures exhibit activity-driven interfacial dynamics. The effective interfacial stiffness, \( \gamma_{\rm eff} \), is modified by activity \( \zeta \):
\begin{equation}
    \gamma_{\rm eff} \approx \gamma_0 -\lambda \zeta
\end{equation}
where \( \gamma_0 \) is the passive surface tension, and \( \lambda \) is a phenomenological coefficient. In extensile systems, active stresses deform interfaces by reducing the effective interfacial stiffness and promote fluctuations that affect the shape and dynamics of droplets (see Fig. \ref{fig:droplets}). If the activity is strong enough, the interface becomes unstable, leading to spontaneous emulsification, where nematic and isotropic domains continuously form and dissolve (see Fig. \ref{fig:emulsion}) \cite{Dogic2022,Dogic2024}.

Active anchoring describes the spontaneous alignment of the nematic director at an active-passive interface due to activity-driven effects. Unlike passive nematics, where anchoring is determined by molecular interactions, active systems are dominated by non-equilibrium effects \cite{Yeomans2014,Coelho2021ptrsa}. The active force at the interface is given by:
\begin{equation}
    \mathbf{F}_{\text{active}} = - \nabla \cdot \Pi^{\text{active}} = \zeta \nabla \cdot Q
\end{equation}
At the interface of an extensile active system, the normal component of the active force points outward from the nematic region when homeotropic anchoring is present, and inward under planar anchoring. As a result, an active droplet is stretched along the director and compressed perpendicular to it (see Fig. \ref{fig:active} panel (b) top row).
The tangential component of the active force generates shear within the nematic, which in turn exerts a torque that reorients the director. Since the direction of shear depends on the director's orientation, this feedback stabilizes planar alignment in extensile systems (see Fig. \ref{fig:active} panel (b) bottom row) \cite{Yeomans2014,Coelho2021ptrsa}. In contractile systems, the active forces act in a similar but symmetrical manner, leading to opposite tendencies in deformation and alignment \cite{Yeomans2014,Coelho2021ptrsa}.  Activity-driven effects can induce anchoring transitions and novel interfacial behaviors, including active wetting transitions \cite{Dogic2022,Coelho2023}, where the apparent contact angle of droplets on surfaces vanishes.

Active nematic emulsions behave fundamentally differently from passive emulsions. Unlike passive emulsions with stable interfaces, active nematic droplets move and deform dynamically, exhibiting oscillatory or chaotic persistent motion  (see Fig. \ref{fig:droplets}) \cite{Marchetti2022,Coelho2023,Giomi2014}. Topological defects within active droplets drive collective behaviors such as droplet fusion or division \cite{Coelho2023,Giomi2014}. Activity can also induce transitions from static interfaces to self-propelled front-like structures \cite{Coelho2019,Coelho2020,Coelho2024} and influence defect motion at interfaces, leading to transient structures or steady-state patterns \cite{Doostmohammadi2024,Coelho2022}.
At low activity, stable phase-separated domains may persist due to the interplay between surface tension and active stresses (see Fig. \ref{fig:emulsion} panel (a)). 
These phenomena remain an active area of research, with theoretical, numerical, and experimental studies continuously advancing our understanding of active matter.

\section{Conclusion}

In this work, we presented a semi-analytical yet comprehensive framework to describe phase separation and interfacial phenomena in mixtures of nematic and isotropic fluids. By coupling the Cahn--Hilliard model, which governs concentration fluctuations, with the Landau--de Gennes theory for nematic ordering, and incorporating hydrodynamic equations for fluid flow, we derived a unified set of governing equations that captures the essential physics of these complex systems. The inclusion of activity, which drives the system out of equilibrium, further enriches the phenomenology—giving rise to spontaneous flows, active defect dynamics, and the emergence of dynamic steady states.

The phase separation dynamics of nematic--isotropic mixtures display a diverse range of behaviors, resulting from the interplay between thermodynamic driving forces, kinetic constraints, and the coupled evolution of conserved and non-conserved order parameters. At equilibrium, interfacial properties are shaped by elastic anisotropy, anchoring conditions, and fluctuation-induced effects. In active systems, these are further modified by internally generated stresses, leading to novel features such as active emulsification, and spontaneous interfacial instabilities.

While the theoretical and computational models discussed here successfully reproduce many qualitative and quantitative aspects of both passive and active nematic--isotropic mixtures, several open challenges remain. Bridging the gap between microscopic interactions and macroscopic emergent behavior continues to be a central goal, particularly in systems with complex molecular architecture or under confinement. A key avenue for improvement lies in the more accurate incorporation of anisotropic elasticity into both hydrodynamic and interfacial models, which will refine our understanding of flow-induced alignment, defect creation and annihilation, and the mechanical response of curved interfaces.

Advancements in numerical methods, multiscale modeling, and high-resolution experiments are expected to further deepen our insight into these rich systems. Looking ahead, future research will increasingly focus on non-uniform and geometrically constrained environments—such as patterned substrates, curved geometries, and microfluidic channels—where confinement and topology introduce new effects. Additionally, coupling to external fields (e.g., electric, magnetic, or optical) offers promising strategies for dynamically tuning phase behavior and controlling interfacial structures. Together, these directions will advance both fundamental understanding and practical applications of nematic fluids in soft matter and materials science.

\section*{Acknowledgements}
We acknowledge financial support from the Portuguese Foundation for Science and Technology (FCT) under the project UID/00618/2025, DL 57/2016/CP1479/CT0057 (DOI 10.54499/DL57/2016/CP1479/CT0057) and 2023.10412.CPCA.A2 (DOI 10.54499/2023.10412.CPCA.A2). 

\bibliographystyle{ar-style4}
\bibliography{refs} 

\end{document}